\documentclass[twocolumn,showpacs,floats,floatfix,superscriptaddress,aps,pra]{revtex4-1}
\usepackage{amsfonts}
\usepackage{amssymb}
\usepackage{amsmath}
\usepackage{calc}
\usepackage{graphicx}
\usepackage{bm}

\usepackage[normalem]{ulem}
\usepackage{amsmath,amssymb}
\usepackage{multirow}
\usepackage{xcolor,soul}
\usepackage{srcltx}
\usepackage{hyperref,graphicx}

\def\be{ \begin{equation}}
\def\ee{ \end{equation}}
\def\bea{ \begin{eqnarray}}
\def\eea{ \end{eqnarray}}
\def\bse{ \begin{subequations}}
\def\ese{ \end{subequations}}
\def\bc{ \begin{center}}
\def\ec{ \end{center}}

\begin{document}

\author{Stefano Longhi$^{*}$} 
\affiliation{Dipartimento di Fisica, Politecnico di Milano and Istituto di Fotonica e Nanotecnologie del Consiglio Nazionale delle Ricerche, Piazza L. da Vinci 32, I-20133 Milano, Italy}
\email{stefano.longhi@polimi.it}

\title{Topological pumping of edge states via adiabatic passage}
  \normalsize


%
\bigskip
\begin{abstract}
\noindent  
Topological pumping of edge states in finite crystals or quasicrystals with non-trivial topological phases provides a powerful means for robust excitation transfer. In most schemes of topological pumping, the edge states become delocalized and immersed into the continuum during the adiabatic cycle, requiring extremely slow evolution to avoid nonadiabatic effects. Here a scheme of topological pumping based on adiabatic passage of edge and interface states is proposed, which is more robust to nonadiabatic effects and avoids delocalization of the state over the entire adiabatic cycle. The scheme is illustrated by considering adiabatic passage in a system sustaining two topologically-protected edge states and one interface state, realized by interfacing two dimerized Su-Schrieffer-Heeger chains with different topological order.
\end{abstract}



\maketitle

\section{Introduction}
Thouless pumping \cite{R1}, i.e. the quantized transport in a one-dimensional cyclically modulated
periodic potential, is a cornerstone of condensed matter physics and provides a fundamental example of topology in quantum systems \cite{R2,R3,R3bis}. 
 Experimental demonstrations of
Thouless pumping have been reported in a wide variety of physical systems,
including few-body semiconductor quantum dots \cite{R4,R5},  ultra-cold atoms in optical lattices \cite{R6,R7,R8,R9}, photons in optical waveguide lattices \cite{R10,R11,R12}, and artificial spin systems \cite{R13,R14}.
 Pumping occurs when system parameters are varied in a
cyclic manner and sufficiently slowly that the quantum system always remains in its ground state.
The cyclic adiabatic pumping shows an intrinsic geometric character and the charge transport over one cycle is related to the Chern number, resulting in the quantization of the excitation transport \cite{R1,R3}. Having a
topological nature, the transport scheme turns out to be robust against disorder, making  topological pumping particularly relevant in applications such as electronic transport in mesoscopic structures, quantum state transfer, quantum entanglement, quantum information processing, etc. \par
Interestingly, in finite systems with open boundaries sustaining topologically-protected edge states, topological pumping can be exploited to realize robust 
excitation transfer by adiabatic evolution of edge states \cite{R3bis,R10,R11,R15,R16}. This effect is connected to the appearance, in the
Hamiltonian spectrum, of gapless points enclosed by the
adiabatic loop in parameter space \cite{R3,R3bis}. Adiabatic pumping of edge states goes beyond perfect crystal systems and can be observed for quasicrystals \cite{R10,R11} and for non-Hermitian crystals \cite{R16}. Other methods for robust adiabatic transport of topological edge states have been reported, which do not require cyclic evolution of the Hamiltonian (see, for instance, \cite{R18,R19,R20,R21,R22}). 
Perhaps, the most studied one-dimensional model which enables topological pumping is the Su-Schrieffer-Heeger (SSH)
model of polyacetylene \cite{referee1,referee2}. This model entails a bipartite tight-binding chain for nearly-free $\pi$ electrons, whose dynamics is coupled to the lattice degrees of freedom (i.e. deformations of the chain), displaying soliton-like localized excitations at domain walls, i.e. at inhomogeneities in
the dimerization pattern of the chain \cite{referee1,referee2}. Such solitons play a major role in charge-transfer doping mechanism and in the conduction properties of semiconductor polymers \cite{referee2,referee3,referee4}.  The coupling between electronic and structural excitations and the related charge transport can be described analytically in the continuous limit of the SSH model \cite{referee5}, where simple expression of drifting solitons at moving dimerization inhomogeneities can be obtained. On the other hand, in the most recent studies dealing with topological properties of the SSH model \cite{R3,R3bis} the $^{\prime}$electronic$^{\prime}$ excitations in the chain are assumed independent of the structural (phononic) degrees of freedom, and inhomogeneities in the dimerization pattern, sustaining topologically-protected localized states, are introduced as structural defects. In this case robust excitation transport arises from external adiabatic change of the hopping rates in the dimerization pattern of the chain (see e.g. \cite{R21}) according to the original idea of Thouless pumping.\\
 A different yet important protocol of robust excitation transfer that has received an increasing interest in recent years is provided by coherent tunneling by adiabatic passage (CTAP) \cite{R23,R24,R25,R26,R27,R27bis,R28,R29,R30,R30a,R30b,R30c}. CTAP can be regarded as the spatial analogue of stimulated Raman adiabatic passage, earlier introduced in atomic ad molecular physics \cite{R31,R31b,R32}. Like Thouless pumping, the transport in CTAP is of geometric nature and thus shows topological protection \cite{R33}. The multi-level extension of CTAP \cite{R30,R32,R34,R35} can realize, in principle, topological transport of excitation between edge sites of long chains in a fashion which is similar to the topological pumping scheme introduced in more recent works \cite{R21}.\\ 
Like any adiabatic method, Thouless pumping of edge states requires strict adiabatic evolution of the Hamiltonian. In fact, despite its topological nature, Thouless pumping is not robust to nonadiabatic effects, which are revealed by non-quantized charge transport \cite{R4,R36}. In topological pumping of edge states, nonadiabatic effects result in degradation of transfer efficiency (or fidelity in quantum state transfer problems). Extremely slow evolution is thus required to realize excitation transfer in  long chains, where over one modulation cycle the topological edge states delocalize and get immersed or touch the continuum bands \cite{R3bis,R10}. Similar restrictions arise in multi-state CTAP protocols.\par
In this work we combine the concepts of topological pumping of edge states and CTAP, suggesting a scheme for efficient topological transfer which is more robust to nonadiabatic effects as compared to Thouless pumping and multi-state CTAP. The main idea is illustrated by considering CTAP among topologically-protected edge and interface states obtained by interfacing two dimerized  SSH chains with different topological order \cite{R37}. Contrary to other topological pumping schemes, such as Thouless pumping in the Rice-Mele model \cite{R3,R3bis,R8}, quasicrystals \cite{R11}, or topological pumping in the dimerized SSH chain \cite{R21}, in the present protocol the energies of localized states remain in the gap and do not enter in the continuum, i.e. edge and interface states remain localized during the entire adiabatic cycle, making the adiabatic restriction less stringent and the transport process faster.

\section{Coherent tunneling by adiabatic passage of topological states}
\subsection{Model}
We consider a tight-binding chain with open boundary conditions comprising an odd number $(2N-1)$ of sites, with $N$ even, obtained by interfacing two SSH chains with different topological order \cite{R37}, as schematically shown in Fig.1(a). The second-quantization Hamiltonian of the system reads
\begin{equation}
\hat{H}=\sum_{n,m} \hat{c}^{\dag}_n H_{n,m} \hat{c}_m
\end{equation}
where $\hat{c}^{\dag}_n$ is the creation operator of a particle at site $n$ ($n=1,2,....,2N-1$) and $H$ is the $(2N-1) \times (2N-1)$ matrix of hopping amplitudes in the Wannier basis. In the nearest-neighbor approximation,  
the matrix $H$ takes the form
\begin{equation}
H_{n,m}=(H_0)_{n,m}+t_1^{\prime}( \delta_{n,N} \delta_{m,N+1}+\delta_{n,N+1} \delta_{m,N}).
\end{equation}
In the above equation, $H_0$ is the Hamiltonian of the uncoupled SSH chains, namely
\begin{figure}[htbp]
  \includegraphics[width=86mm]{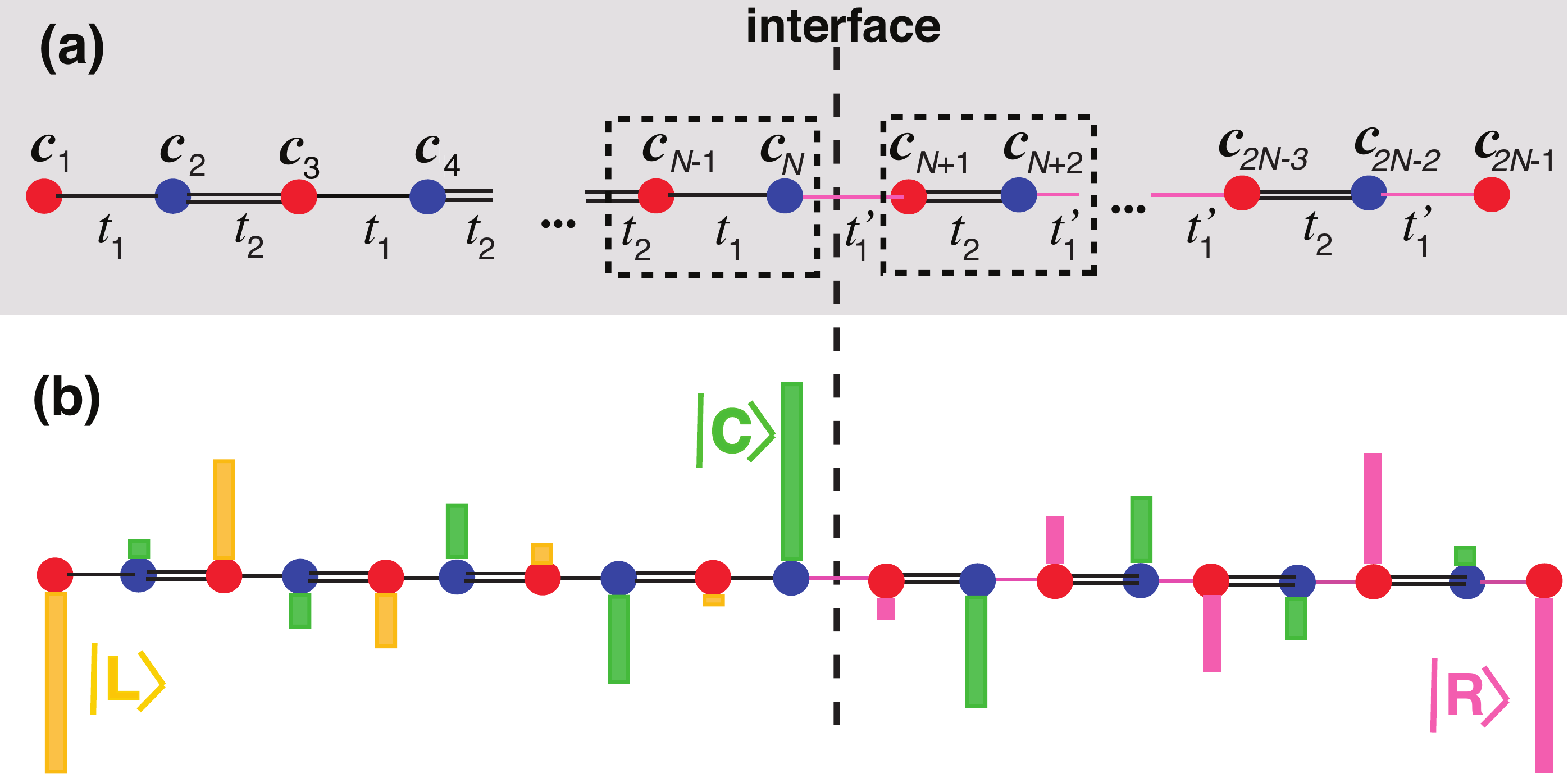}\\
   \caption{(color online) (a) Schematic of the tight-binding chain with an interface connecting two SSH chains with different topological phases. The hopping rates $t_1$ and $t_1^{\prime}$  are smaller than the hopping rate $t_2$, so that in the thermodynamic limit the chain sustains three zero-energy topologically-protected bound states. (b) Schematic of the amplitude distributions of the three nearly-degenerate zero-energy topological modes (left-edge state $|L \rangle$, interface state $|C \rangle$, and right-edge state $|R \rangle$). In the flat band limit $t_1, t_1^{\prime} \rightarrow 0$, the three states correspond to single-site occupation at sites $n=1$ (for $|L \rangle$), $n=N$ (for $|C \rangle$), and $n=2N-1$ (for $|R \rangle$).}
\end{figure}
\begin{equation}
H_0=
\left(\begin{array}{c|c} H_1 & 0 \\\hline 0 & H_2 \end{array}\right)
\end{equation}
where $H_1$ and $H_2$ are two $N \times N$ and $(N-1) \times (N-1)$ matrices, respectively, given by
\begin{equation}
H_1=\left(
\begin{array}{cccccccccc}
0 & t_1 & 0 & 0 & 0 & ... & 0 & 0 & 0 & 0\\
t_1 & 0 & t_2 & 0 & 0 &... & 0 & 0 & 0 & 0 \\
0 & t_2 & 0 & t_1 & 0 & ... & 0 & 0 & 0 & 0\\
0 & 0 & t_1 & 0 & t_2 & ... & 0 & 0 & 0 & 0\\
...& ... & ... & ... & ...& ... & ... & ... & ... & ...\\
0 & 0 & 0 & 0 &0 & ... & t_1 & 0 &t_2 &0 \\
0 & 0 & 0 & 0 &0 & ... & 0 & t_2 &0 &t_1\\
0 & 0 & 0 & 0 &0 & ... & 0 & 0 &t_1 &0
\end{array}
\right).
\end{equation}
and
\begin{equation}
H_2=\left(
\begin{array}{cccccccccc}
0 & t_2 & 0 & 0 & 0 & ... & 0 & 0 & 0 & 0\\
t_2 & 0 & t_1^{\prime} & 0 & 0 &... & 0 & 0 & 0 & 0 \\
0 & t_1^{\prime} & 0 & t_2 & 0 & ... & 0 & 0 & 0 & 0\\
0 & 0 & t_2 & 0 & t_1^{\prime} & ... & 0 & 0 & 0 & 0\\
...& ... & ... & ... & ...& ... & ... & ... & ... & ...\\
0 & 0 & 0 & 0 &0 & ... & t_1^{\prime} & 0 &t_2 &0 \\
0 & 0 & 0 & 0 &0 & ... & 0 & t_2 &0 &t_1^{\prime} \\
0 & 0 & 0 & 0 &0 & ... & 0 & 0 &t_1^{\prime} &0
\end{array}
\right). .
\end{equation}
The parameters $t_1$, $t_2$, $t_1^{\prime}$ entering in the above equations are the alternating hopping rates as shown in Fig.1(b). We assume $t_1, t_1^{\prime}<t_2$, so that the chain sustains a topologically-protected interface state $| C \rangle$ \cite{R37}. In addition, for open boundary conditions there are two additional edge states, localized at the left and right sides of the chain, which we denote by $| L \rangle$ and $| R \rangle$, respectively.  In the thermodynamic limit ($N$ large), the three bound states $|L \rangle$, $| C \rangle$ and $|R \rangle$ are nearly degenerate zero-energy modes, with a typical distribution of excitation which is schematically depicted in Fig.1(b). The three states are defined by
\begin{widetext}
\begin{equation}
|L \rangle= \mathcal{N}_L \left(
\begin{array}{c}
1 \\
0 \\
X \\
0 \\
X^2 \\
0 \\
X^3 \\
0 \\
X ^4 \\
0 \\
... \\
...
\end{array}
\right) \; , \; 
|R \rangle= \mathcal{N}_R \left(
\begin{array}{c}
... \\
... \\
0 \\
Y^4 \\
0\\
Y^3 \\
0 \\
Y^2 \\
0 \\
Y \\
0  \\
1
\end{array}
\right) \; , \;
|C \rangle =
\mathcal{N}_C \left(
\begin{array}{c}
... \\
X^3 \\
0 \\
X^2 \\
0 \\
X \\
1 \\
Y\\
0 \\
Y^ 2 \\
0 \\
Y^3 \\
...
\end{array}
\right)
\end{equation}
\end{widetext}
where 
\begin{equation}
X \equiv -t_1/t_2 \;, \; \; Y \equiv -t_1^{\prime} /t_2
\end{equation}
 and $\mathcal{N}_{L,R,C}$ are normalization constants, given by
\begin{equation}
\mathcal{N}_L  =  \sqrt{\frac{X^2-1}{X^{N}-1}}  
\end{equation}
\begin{equation}
\mathcal{N}_R  =  \sqrt{\frac{Y^2-1}{Y^{N}-1}} 
\end{equation}
\begin{equation}
\mathcal{N}_C  =  \left( {\frac{X^{N}-1}{X^2-1}+\frac{Y^{N}-1}{Y^2-1}-1} \right)^{-1/2}. 
 \end{equation}
 Such states are exact eigenvectors of $H$ with zero energy only in the large $N$ limit; for small $N$ such states ibridize and energy degeneracy is lifted, as discussed in the next subsection [Fig.2(a)]. Note that in the flat band limit $t_1/t_2 \rightarrow 0$, $t_1^{\prime}/t_2 \rightarrow 0$ (i.e. $X,Y \rightarrow 0$) the three eigenstates $|L \rangle$, $|C \rangle$ and $|R \rangle$ correspond to single-site occupation of sites $n=1$, $n=N$ and $n=2N-1$, respectively.
\subsection{Topological pumping and approximate three-level description}
 The main idea of the transfer scheme is to realize CTAP among the three topological bound states $|L \rangle$, $|C \rangle$ and $|R \rangle$ by adiabatically evolving the system in its dark state. 
 In the spirit of CTAP \cite{R28,R32}, let us assume that the hopping amplitude $t_2$ is constant while the amplitudes $t_1$ and $t_1^{\prime}$ are varied in time as follows
\begin{equation}
t_1= \Omega(t-\delta/2) \;\; ,\; \;\;  t_1^{\prime}= \Omega(t+\delta/2)
\end{equation}
where $\Omega(t)$ is a bell-shaped function with a maximum $\Omega_m<t_2$ at $t=0$ and with $\Omega(t) \rightarrow 0$ as $t \rightarrow \pm \infty$, and $\delta>0$ is a delay constant. A typical instantaneous energy spectrum of $H(t)$, where $t$ is considered as a parameter, is shown in Fig.2(a) for a Gaussian function $\Omega(t)=\Omega_m \exp (-t^2/w^2)$ [Fig.2(b)]. Note that the limits $t \rightarrow \pm \infty$ correspond to the flat band limits $t_1=t_1^{\prime}=0$. Since $t_1$ and $t_1^{\prime}$ remain smaller than $t_2$, the energy gap does not close as $t$ is varied. In the gap, there are three bound states, one with zero energy and the other two with opposite energies emanating from zero as $t \rightarrow \pm \infty$ [Fig.2(c)]. As shown below, such eigenstates of $H$ correspond to suitable linear combinations of the states $|L \rangle$,  $|R \rangle$ and $|C \rangle$ defined by Eq.(6), and the zero-energy eigenstate defines the dark state in CTAP theory. Interestingly, such a zero-energy eigenstate is fully localized at the left edge site $n=1$ of the chain for $t \rightarrow -\infty$, and at the right edge site $n=2N-1$ of the chain as $t \rightarrow \infty$ [see the insets of Fig.2(c)]. This means that topological pumping of excitation from left to right edge sites of the chain can be realized by adiabatic evolution of $H(t)$ in the zero-energy (dark) eigenstate.\\
\begin{figure*}[htbp]
  \includegraphics[width=178mm]{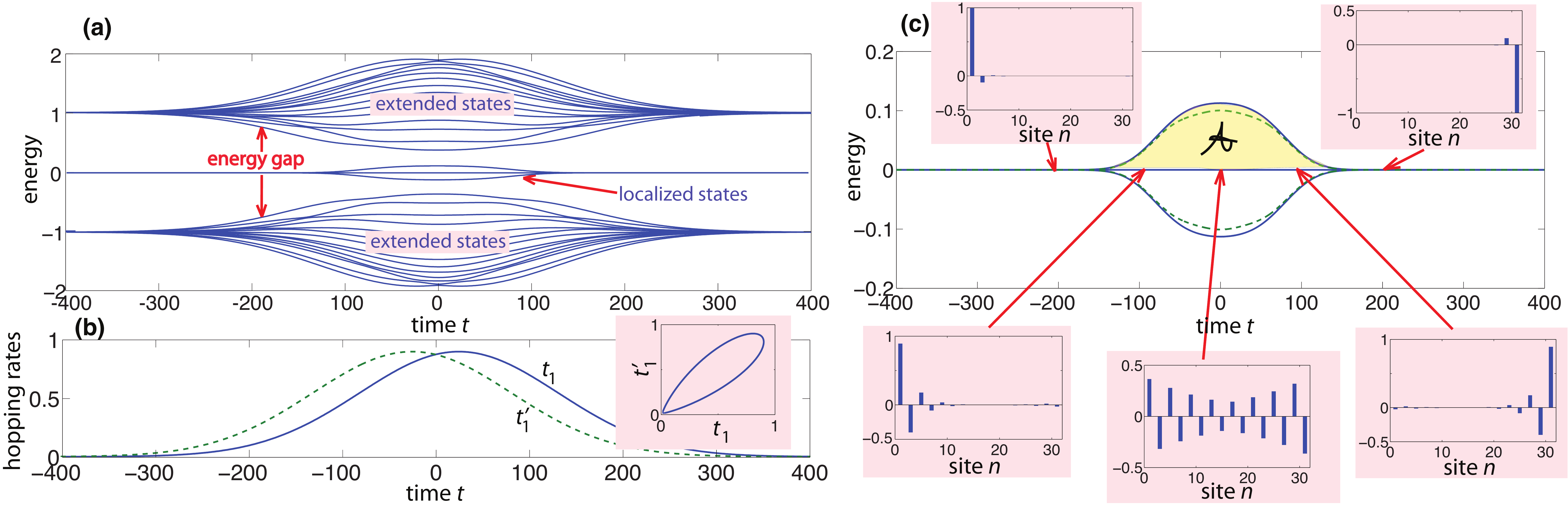}\\
   \caption{(color online) (a) Numerically-computed instantaneous energy spectrum of $H(t)$ for $t_1=\Omega(t-\delta/2)$, $t_1^{\prime}=\Omega(t+\delta/2)$, with $\Omega(t)=\Omega_m \exp(-t^2/w^2)$. Parameter values are $t_2=1$, $\Omega_m=0.9$, $w=150$ and $\delta=w/3=50$. The chain comprises $2N-1=31$ sites. The energy spectrum consists of two bands of delocalized (Bloch) states separated by an energy gap, and three localized states with energies in the gap. (b) Behavior of the hopping amplitudes $t_1$ and $t_1^{\prime}$ versus time $t$. The inset shows the trajectory in the $(t_1, t_1^{\prime})$ plane. (c) Detailed behavior of the energies of the three localized modes versus $t$ (solid curves). The dashed curves depict the energies of the localized modes as predicted by the three-mode CTAP approximation [Eq.(16)]. The insets show the amplitude distributions of the zero-energy localized state (dark state) for a few increasing values of $t$ ($t=-200,-100,0,100,200$). Adiabatic evolution of the zero-energy topological state is ensured provided that the area $\mathcal{A}$, indicated by the shaded area, is much larger than $ \pi/2$.}
\end{figure*}
To show the connection between such a topological pumping scheme and CTAP, let us consider the dynamical evolution of the system in the subspace described by the vectors $|L \rangle$, $|C \rangle$ and $|R \rangle$, defined by Eq.(6).  After expanding the state vector as 
\begin{equation}
| \psi(t) \rangle \simeq a_L(t) |L \rangle +a_C(t) |C \rangle +a_R(t) |R \rangle
\end{equation}
from the Schr\"odinger equation $i \partial_t | \psi(t) \rangle=H(t) | \psi(t) \rangle$ the following reduced three-level equations for the evolution of the amplitudes $a_{L,C,R,} (t)$ can be obtained (see Appendix A) 
\begin{equation}
i \frac{d}{dt}
\left( 
\begin{array}{c}
a_L \\
a_C \\
a_R
\end{array}
\right)=\left(
\begin{array}{ccc}
0 & \Omega_L & 0 \\
\Omega_L & 0 & \Omega_R \\
0 & \Omega_R & 0
\end{array}
\right)\left( 
\begin{array}{c}
a_L \\
a_C \\
a_R
\end{array}
\right) \equiv H_{red} \left( 
\begin{array}{c}
a_L \\
a_C \\
a_R
\end{array}
\right)
 \end{equation}
 where we have set
 \begin{eqnarray}
 \Omega_L(t) & \equiv & \langle L | H |C \rangle   =  \mathcal{N}_L \mathcal{N}_C t_1 X^{(N/2-1)} \\
 \Omega_R(t) & \equiv & \langle R | H |C \rangle  =  \mathcal{N}_R \mathcal{N}_C t_1^{\prime} Y^{(N/2-1)} .
 \end{eqnarray}
Clearly, Eqs.(13) are the  basic equations of CTAP in a three-state system \cite{R28,R32}. The instantaneous eigenvalues $E_0, E_{\pm}$ and corresponding eigenvectors $|\phi_0 \rangle$, 
$| \phi_{\pm} \rangle$ of the reduced matrix $H_{red}$ entering in the three-state dynamics [Eq.(13)] are given by
\begin{equation}
E_0=0 \; , \;\; E_{\pm}= \sqrt{\Omega_L^2+\Omega_R^2}
\end{equation}
and
\begin{equation}
|\phi_0 \rangle= 
\left(
\begin{array}{c}
\frac{\Omega_R}{\sqrt{\Omega_L^2+\Omega_R^2}} \\
0 \\
-\frac{\Omega_L}{\sqrt{\Omega_L^2+\Omega_R^2}} 
\end{array}
\right) , \; 
|\phi_{\pm} \rangle= 
\left(
\begin{array}{c}
\frac{\Omega_L}{\sqrt{\Omega_L^2+\Omega_R^2+E_{\pm}^2}} \\
 \frac{E_{\pm}}{\sqrt{\Omega_L^2+\Omega_R^2+E_{\pm}^2}} \\
\frac{\Omega_R}{\sqrt{\Omega_L^2+\Omega_R^2+E_{\pm}^2}} 
\end{array}
\right).
\end{equation}
The zero-energy eigenstate $|\phi_0 \rangle$ is referred to as the dark state since in this state the topological interface mode $|C \rangle$ is not excited, contrary to the other two eigenstates $| \phi_{\pm} \rangle$.
 For the time-dependence of hopping amplitudes $t_1, t_1^{\prime}$ defined by Eq.(11), from Eqs.(14) and (15) it readily follows that $\Omega_{L}(t) / \Omega_R(t) \rightarrow 0$ as $t \rightarrow -\infty$ and $\Omega_{R}(t) / \Omega_L(t) \rightarrow 0$ as $t \rightarrow -\infty$, so that the dark state adiabatically evolves from $|\phi_0 \rangle=(1,0,0)^T$ as $t \rightarrow -\infty$, to  $|\phi_0 \rangle=(0,0,-1)^T$ as $t \rightarrow \infty$. The condition of global adiabaticity is expressed by the area condition \cite{R32}
 \begin{equation}
 \mathcal{A} \equiv \int_{-\infty}^{\infty} dt  \sqrt{\Omega_L(t) ^2+\Omega_R(t) ^2} \gg \pi/2.
 \end{equation}
 Note that the area $\mathcal{A}$ corresponds to the shaded area in the adiabatic diagram of Fig.2(c). Clearly, owing to the exponential dependence of $\Omega_{L,R}$ on $N$ [Eqs.(14) and (15)], longer interaction times are required to ensure adiabatic following as the number of sites in the chain is increased.  \\
 An example of topological pumping for a SSH chain comprising $(2N-1)=31$ sites is shown in Fig.3. The values of parameters and the time variation of the hopping amplitudes $t_1$ and $t_1^{\prime}$ are the same as in Fig.2. At initial time the system is prepared at the left edge site $n=1$, and the evolution of the occupation probabilities $P_1(t)$ and $P_{2N-1}(t)$ of excitation at the left and right edge sites is exactly computed by numerically solving the  Schr\"odinger equation $i \partial_t | \psi(t) \rangle=H(t) | \psi(t) \rangle$ for an interaction time $T=800$. The inset in Fig.3 shows the corresponding evolution of the occupation probabilities $|a_{L,R}(t)|^2$ of instantaneous states $|L \rangle$ and $| R \rangle$ as computed by the approximate three-level CTAP equations (13). To estimate nonadiabatic effects, we numerically computed the transfer probability $P_{2N-1}(T)$ versus the interaction time $T$. The results are summarized in Fig.4. In the numerical simulations we assumed a Gaussian function $\Omega(t)=\Omega_{m} \exp(-t^2/w^2)$ with $w=3T/10$ and a delay time $\delta=w/3$; other parameter values are $2N-1=31$, $t_2=1$ and $\Omega_{m}=0.9$. 
\begin{figure}[htbp]
  \includegraphics[width=82mm]{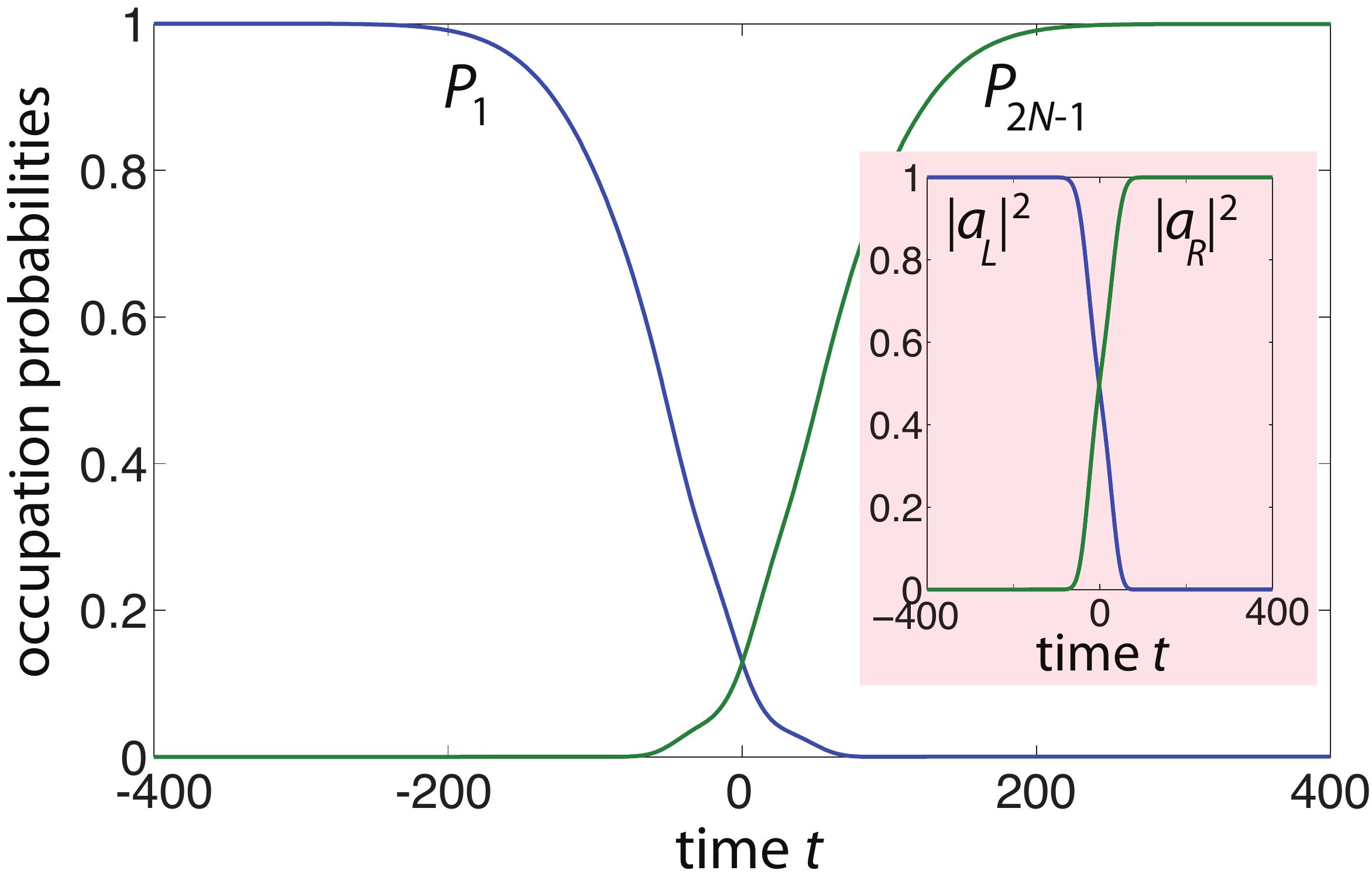}\\
   \caption{(color online) Numerically-computed evolution of the occupation probabilities $P_1$ and $P_{2N-1}$ of left and edges sites in a SSH chain comprising $(2N-1)=31$ sites. Parameter values and the evolution of hopping amplitudes $t_1, t_1^{\prime}$ are as in Fig.2. The inset shows the evolution of the occupation probabilities of edge states $|L \rangle$ and $|R \rangle$ as predicted by the three-level CTAP model. The interaction time is $T=800$.}
\end{figure}
\begin{figure}[htbp]
  \includegraphics[width=82mm]{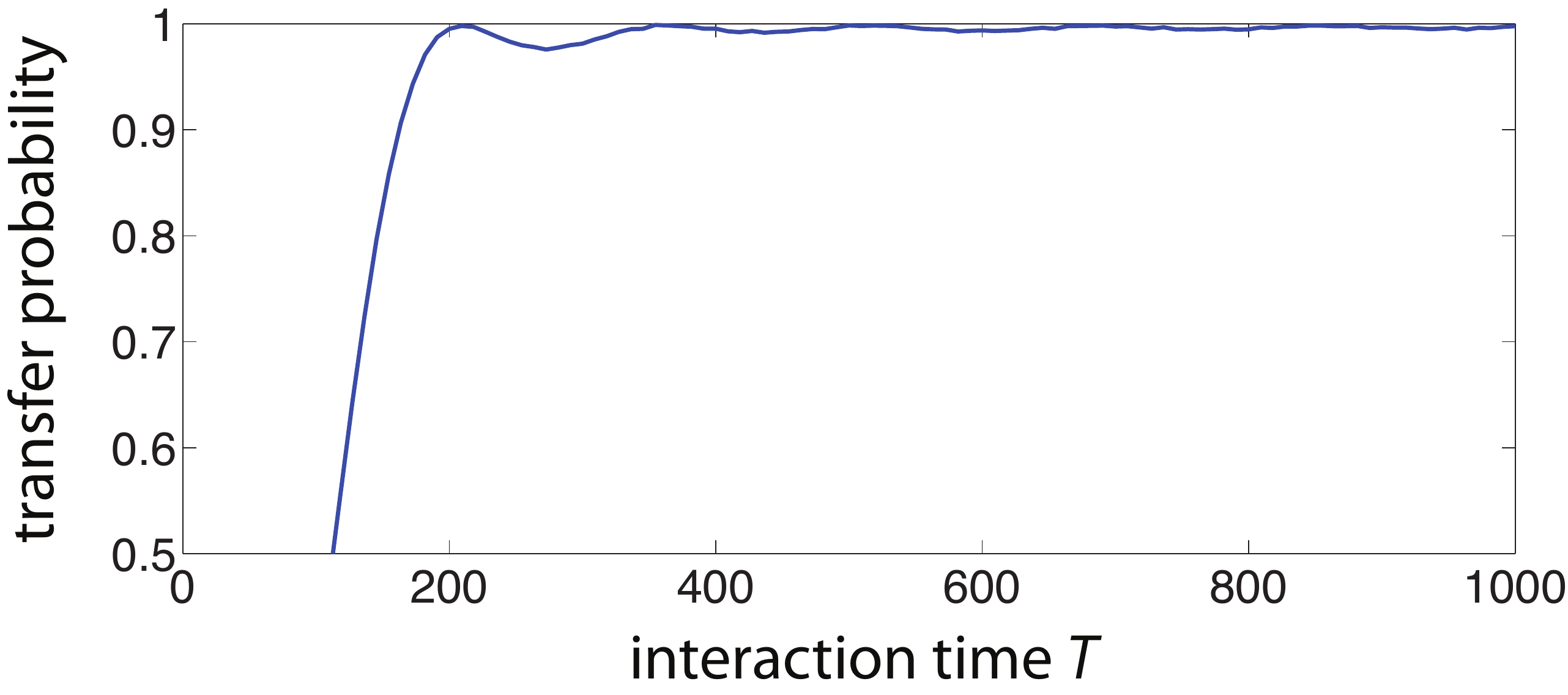}\\
   \caption{(color online) Numerically-computed behavior of the transfer probability versus interaction time. Parameter values are given in the text.}
\end{figure}
\begin{figure}[htbp]
  \includegraphics[width=82mm]{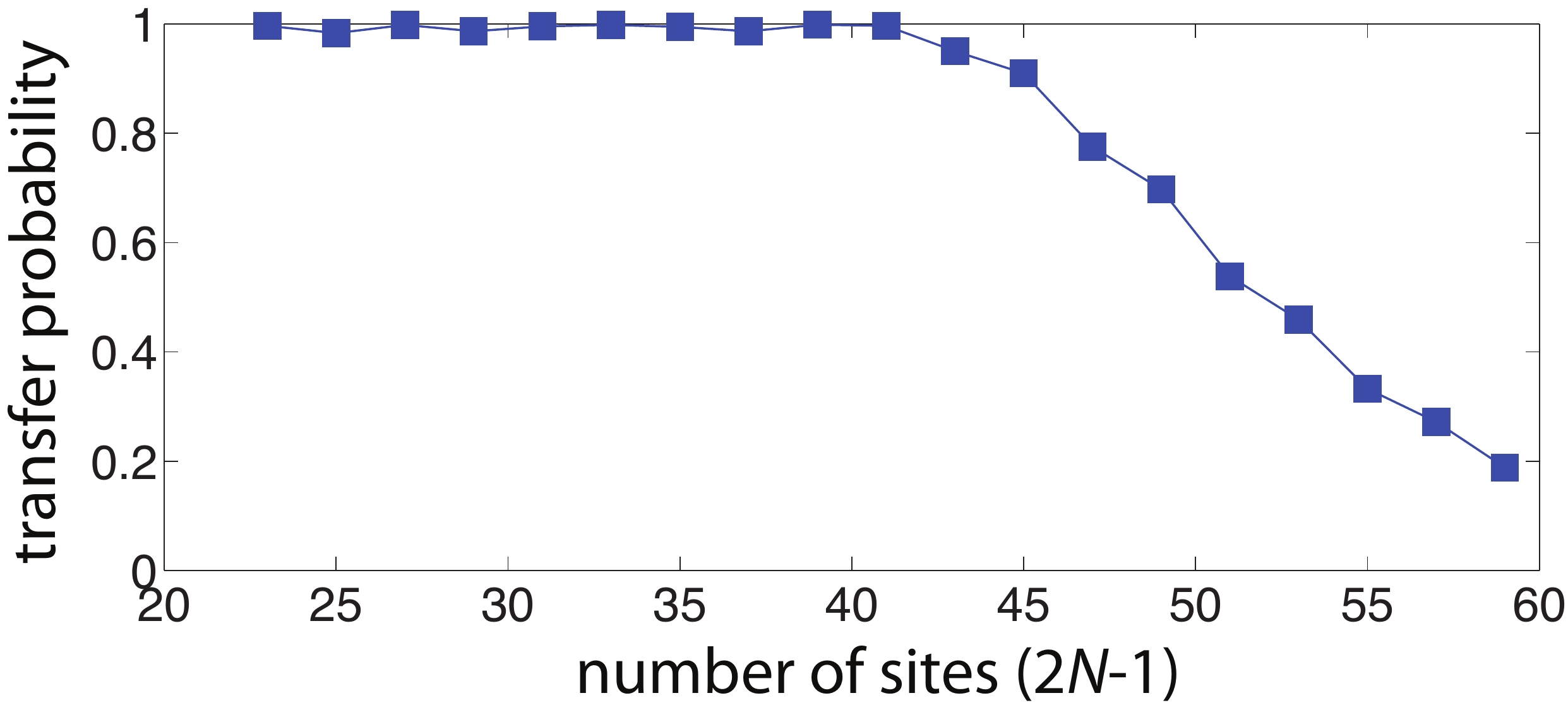}\\
   \caption{(color online) Numerically-computed behavior of the transfer probability versus number of sites in the chain for fixed interaction time $T=400$. Other parameter values are given in the text.}
\end{figure}
 Figure 4 clearly shows that nonadiabatic effects are important for interaction times shorter than $T \sim 160$, where the transfer excitation probability falls below $90 \%$.\\
 For a fixed interaction time, nonadiabatic effects become more pronounced as the number of sites in the chain is increased. The reason thereof is that the area $\mathcal{A}$ is an almost exponentially decreasing function of $N$, and thus the condition for adiabaticity requires a longer interaction time. This is show, as an example, in Fig.5, where the numerically computed behavior of transfer probability is depicted for a few increasing values of the number of sites in the chain. In the simulations the  interaction time is fixed at $T=400$ and other parameters are as in Fig.4 ($w=10 T/3$, $\delta=w/3$, $\Omega_{m}=0.9$, $t_2=1$). Clearly, for a number of sites in the chain larger than $\sim 45 $, nonadiabatic effects become strong and the transfer process highly degraded. 
 \begin{figure*}[htbp]
  \includegraphics[width=178mm]{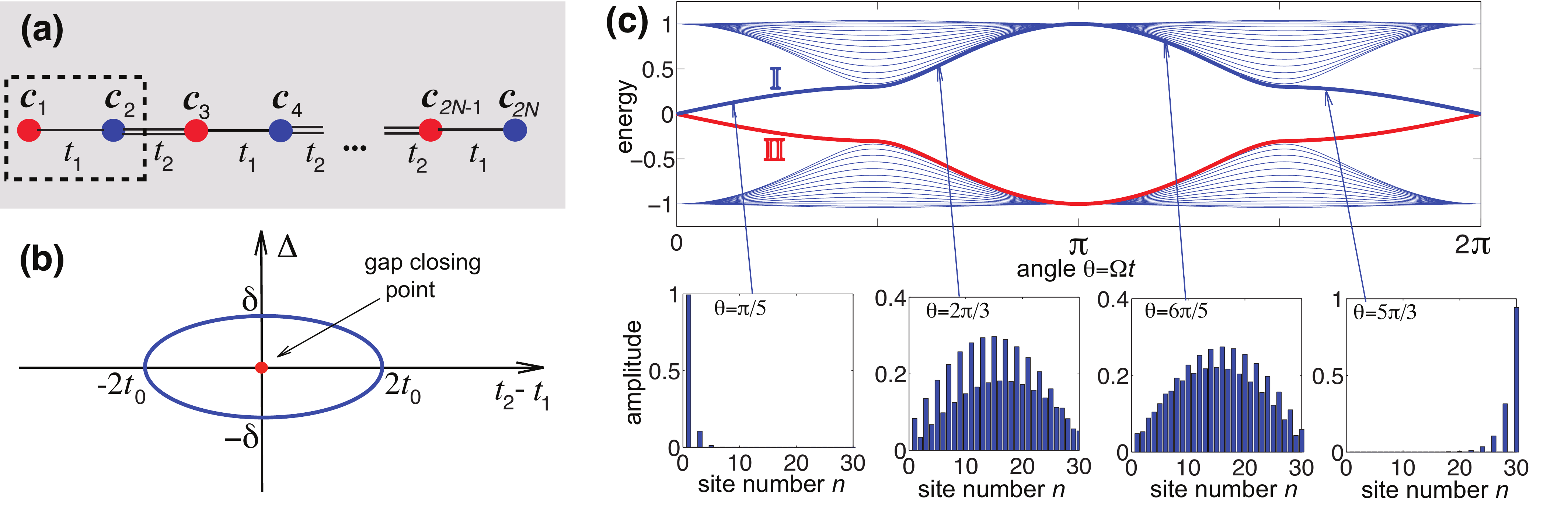}\\
   \caption{(color online) (a) Schematic of the Rice-Mele model with open boundary conditions. A dimerized chain is composed by an integer number $N$ of dimers with intra-dimer hopping amplitude $t_1$ and inter-dimer hopping amplitude $t_2$. Staggered energy potentials $\pm \Delta$ are applied at alternating sites in the chain. (b) Schematic of an evolving Rice-Mele Hamiltonian in the $(t_2-t_1,\Delta)$ plane for the protocol defined by Eq.(20). The closed path is an ellipse that encircles the origin $\Delta=t_2-t_1=0$ (gap closing point). (c) Energy spectrum of $H$ versus the angle $\theta= \Omega t$ over one cycle for the protocol defined by Eq.(20). Parameter values are $t_0=0.5$, $\delta=0.3$. Bold solid lines I and II correspond to left and right edge states at $\theta=0$, respectively. The distributions of amplitudes $|c_n|^2$ of the adiabatically-evolving state labeled by the solid curve I are also shown for a few increasing values of the phase $\theta$.   }
\end{figure*}
\section{Comparison with other topological pumping schemes}
An interesting feature of the topological pumping method proposed in the previous section is that, contrary to Thouless pumping and other topological pumping schemes, the adiabatically evolving zero-energy eigenstate of the Hamiltonian remains in the gap. Such a feature is expected to provide a major robustness against nonadiabatic effects. A sufficient condition for adiabatic evolution is usually expressed by the so-called gap condition, i.e. the Hamiltonian change should be slow as compared to the frequency separation between the adiabatically evolving eigenstate and other instantaneous eigenstates of the Hamiltonian. The gap condition clearly requires that the adiabatic evolving state does not touch nor enter into the continuum of states. Although the gap condition  is only a sufficient condition and  its
failure does not always mean the breakdown of the adiabatic approximation \cite{R38}, it is likely that a wide gap can allow for a  faster and more efficient transfer process. It is thus worth comparing nonadiabatic effects and speed of transfer in our topological pumping scheme with other methods where the gap condition is not met. To this aim, we compare transfer efficiency versus interaction time in a chain of the same size by considering other two  topologically-protected transfer protocols: the Thouless pumping in the Rice-Mele model \cite{R3bis,R16} and the half-cycle topological pumping in the SSH chain \cite{R21}.

\subsection {Thouless pumping of edge states in the Rice-Mele model}
Let us consider a one-dimensional dimerized chain with open boundary conditions comprising an even number $2N$ of sites with alternating hopping amplitudes $t_1$, $t_2$ and staggered potential energies $\pm \Delta$ [Fig.6(a)]. The system is described by the Rice-Mele Hamiltonian \cite{R3bis}, i.e. by Eq.(1) with the following $(2N \times 2N)$ matrix of hopping amplitudes
\begin{equation}
H=\left(
\begin{array}{cccccccccc}
\Delta & t_1 & 0 & 0 &  ... & 0 & 0 & 0 & 0\\
t_1 & -\Delta & t_2 & 0 & ... & 0 & 0 & 0 & 0 \\
0 & t_2 & \Delta & t_1 &  ... & 0 & 0 & 0 & 0\\
0 & 0 & t_1 & -\Delta & t_2  & 0 & 0 & 0 & 0\\
...& ... & ... & ... & ... & ... & ... & ... & ...\\
0 & 0 & 0 & 0  & ... & t_1 & -\Delta &t_2 &0 \\
0 & 0 & 0 & 0 & ... & 0 & t_2 &\Delta &t_1\\
0 & 0 & 0 & 0 & ... & 0 & 0 &t_1 &-\Delta
\end{array}
\right).
\end{equation}
In the thermodynamic limit ($N$ large), the Rice-Mele Hamiltonian shows two energy bands separated by a gap of width $\sqrt{(t_2-t_1)^2+\Delta^2}$, which vanishes at the critical point $\Delta=0$ and $t_1=t_2$. To realize topological pumping of edge states, the hopping amplitudes and potential energy are adiabatically varied along a closed loop that encircles the critical point in the $(t_2-t_1,\Delta)$ plane \cite{R3bis}. For example let us assume
\begin{equation}
t_1=t_0 [1-\cos (\Omega t)], \; t_2=t_0 [1+ \cos (\Omega t)]  , \; \Delta= \delta \sin (\Omega t)
\end{equation}
corresponding to encircling the critical point by an ellipse [Fig.6(b)]. Note that, at $t=0$, one has $t_1=0$ (flat band limit) and $\Delta=0$, so that two zero-energy instantaneous eigenvectors of $H$, denoted by I and II, correspond to single site excitation at the two edge sites (either left or right) of the chain. Starting at $t=0$ with one of the two eigenstates, for example with eigenstate I localized at the left edge, and adiabatically evolving such an eigenstate, after one cycle, i.e. at time $t= 2 \pi / \Omega$, this state is transformed into the right edge state II, thus realizing topological pumping from left to right edge sites of the chain. This is illustrated in the example of Fig.6(c) for a chain comprising $2N=30$ sites. Note that the instantaneous energies of the adiabatically-evolving eigenstates I and II turn out to touch the upper and lower bands, thus becoming fully delocalized states during the adiabatic cycle. Nonadiabatic effects are revealed by plotting the transfer probability $P_{2N}(T)=|c_{2N}(T)|^2$ versus the interaction time $T= 2 \pi / \Omega$, which is obtained by exact numerical solution of the Schr\"odinger equation for the initial condition $c_n(0)=\delta_{n,1}$. The results are shown in Fig.7 for $t_0=0.5$ and for a few values of $\delta$. Note that for such parameter values the largest values taken by $t_1$ and $t_2$ in the oscillation cycle is one, i.e. the same value as in Figs.3 and 4, so that we can safely compare the transfer time and transfer efficiency in the two protocols. A comparison of Figs.4 and 7 clearly shows that nonadiabatic effects are more pronounced in the Thouless pumping scheme (Fig.7) than in the CTAP protocol (Fig.4), transfer probability larger than $90 \%$ requiring in the former case an interaction time larger than $\sim 1600$, i.e. one order of magnitude larger than the one required in the CTAP protocol. Interestingly, in the Thouless pumping protocol for $\delta \rightarrow 0$, i.e. when the ellipse encircling the gap closing point degenerate into a line on the horizontal axis and the gap closes, the transfer probability shows an oscillatory behavior versus the interaction time (see the inset of Fig.7), which is a signature of strong nonadiabaticity. Such an oscillatory behavior stems from Rabi-like oscillations of the edge states, which have been considered in some recent works \cite{R39,R40,R41,R42}.
\subsection{Topological pumping of edge states in the Su-Schrieffer-Heeger chain}
Another adiabatic method for excitation transfer is provided by topological pumping in the SSH chain with an odd number $(2N-1)$ of sites \cite{R21}. Such a method is a slight variation of the multi-state  CTAP considered in earlier works \cite{R28,R32,R34,R35}. The Hamiltonian of the system is defined by Eq.(1) with the following $(2N-1) \times (2N-1)$ hopping amplitude matrix [Fig.8(a)]
\begin{equation}
H=\left(
\begin{array}{cccccccccc}
0 & t_1 & 0 & 0 &  ... & 0 & 0 & 0 & 0\\
t_1 & 0 & t_2 & 0 & ... & 0 & 0 & 0 & 0 \\
0 & t_2 & 0 & t_1 &  ... & 0 & 0 & 0 & 0\\
0 & 0 & t_1 & 0 & t_2  & 0 & 0 & 0 & 0\\
...& ... & ... & ... & ... & ... & ... & ... & ...\\
0 & 0 & 0 & 0  & ... & t_2 & 0 &t_1 &0 \\
0 & 0 & 0 & 0 & ... & 0 & t_1 &0 &t_2\\
0 & 0 & 0 & 0 & ... & 0 & 0 &t_2 & 0
\end{array}
\right).
\end{equation}

\begin{figure}[htbp]
  \includegraphics[width=82mm]{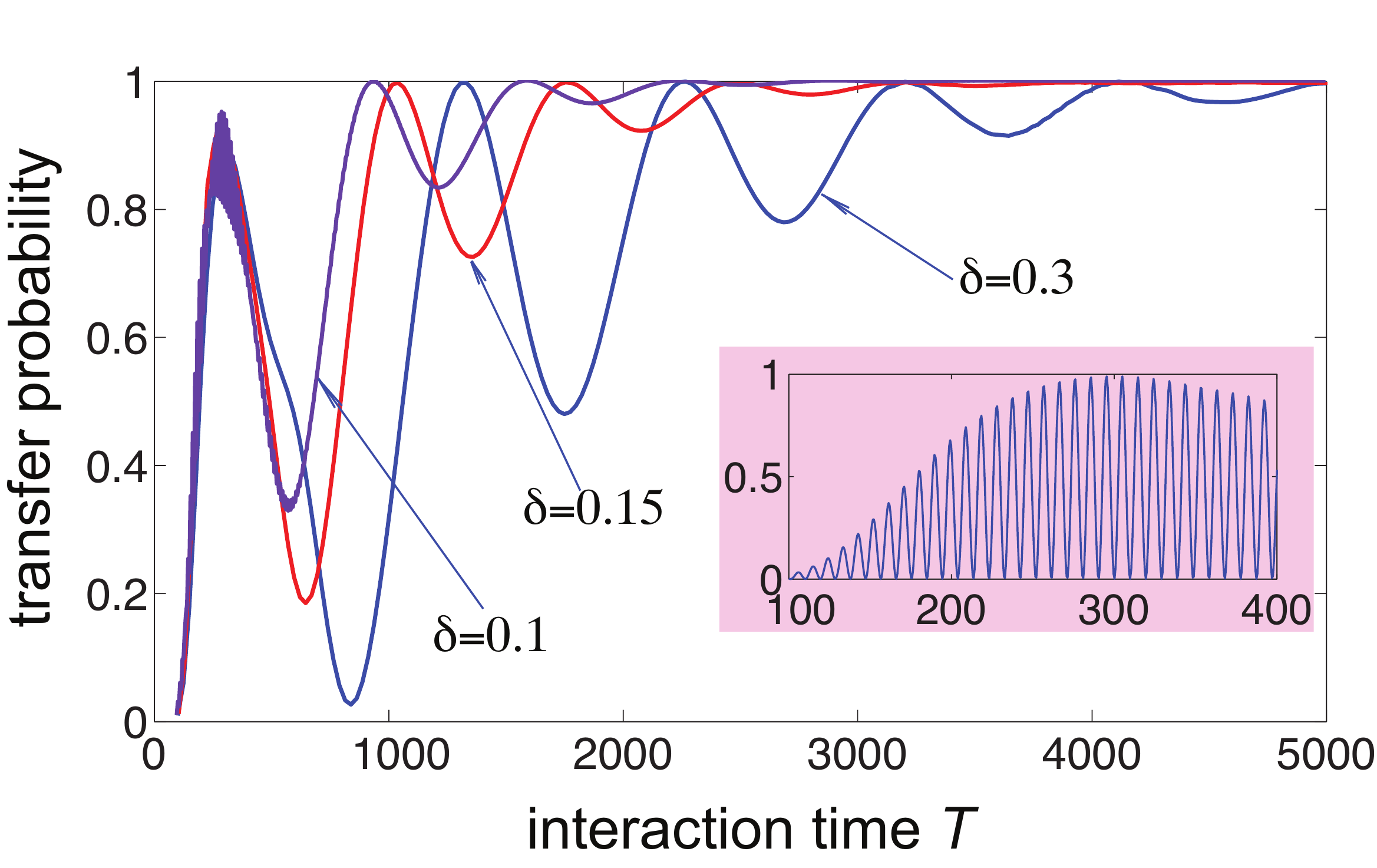}\\
   \caption{(color online) Numerically-computed behavior of the transfer probability versus interaction time in the Rice-Mele chain for a few values of $\delta$. Other parameter values are  $t_0=0.5$ and $2N=30$. The inset shows the behavior of the transfer probability in the limit $\delta=0$. The oscillations correspond to Rabi flopping between zero-energy left and right edge topological states.}
\end{figure}
This Hamiltonian admits of an exact zero-energy eigenstate with site occupation amplitudes $c_n=0$ for $n$ even, $c_n=(-t_1/t_2)^{n-1}$ for $n$ odd ($n=1,2,...,2N-1$). This state is an edge state, localized at either the left or right edges of the chain for $t_1 < t_2$ and $t_1>t_2$, respectively. The localization length diverges as the gap closing point $t_1=t_2$ is attained, where the zero-energy eigenmode is fully delocalized in the chain. To realize topological transfer of edge states, let us consider the following time variation of the hopping amplitudes \cite{R21}
\begin{equation}
t_1(t)=t_0[1-\cos ( \pi t /T)] \; , \; t_2(t)=t_0[1+\cos ( \pi t /T)]
\end{equation}
\begin{figure}[htbp]
  \includegraphics[width=82mm]{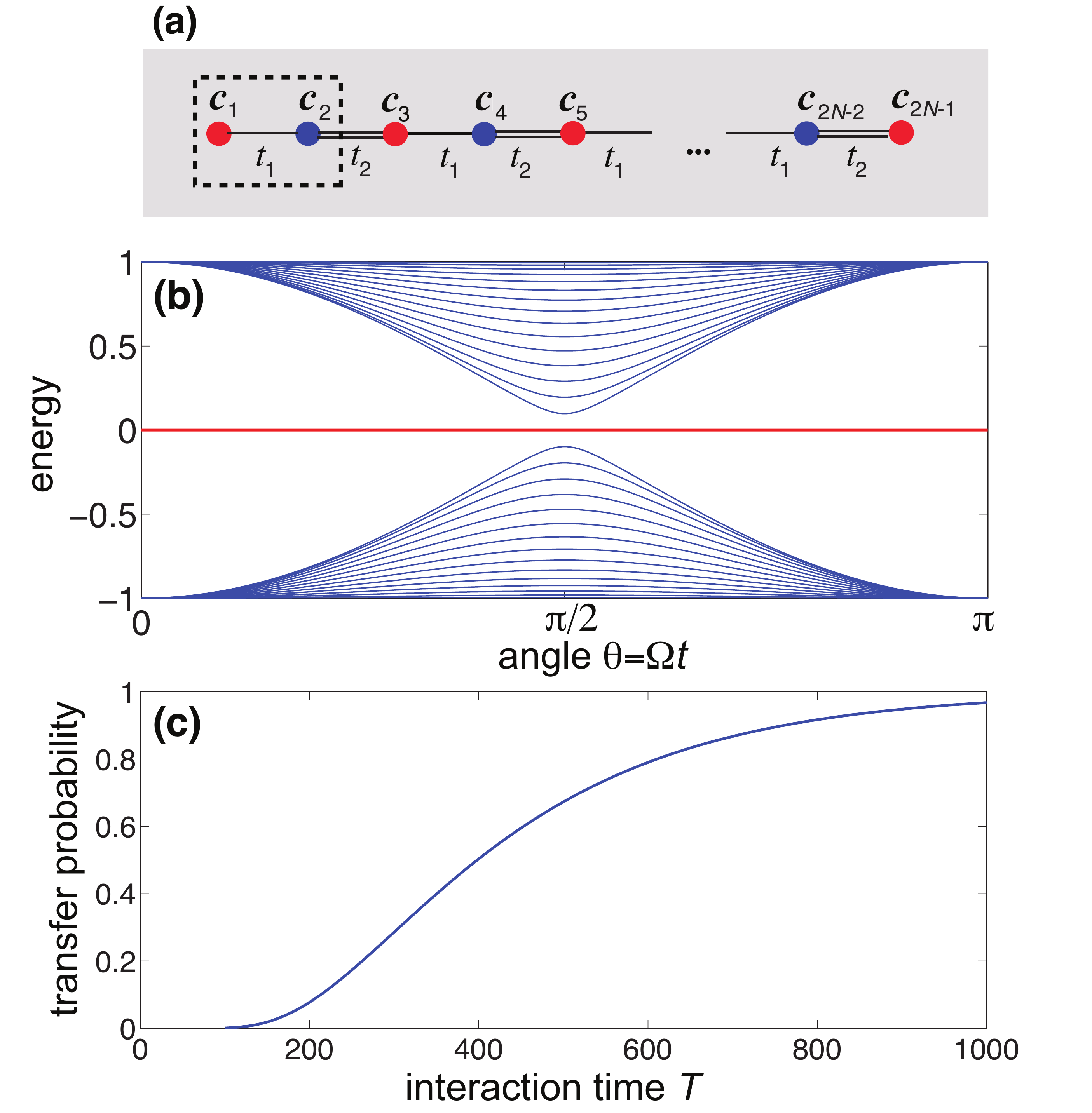}\\
   \caption{(color online) (a) Schematic of a SSH chain with an odd number of sites (half-integer dimers) for adiabatic topological pumping of edge states. (b) Energy spectrum versus the phase $\theta= \pi t /T$ in a chain comprising $(2N-1)=31$ sites for $t_0=0.5$. The bold solid line corresponds to the topologically-protected zero-energy eigenstate. (c) Numerically-computed behavior of the transfer probability versus interaction time $T$. The hopping amplitudes $t_1$ and $t_2$ vary in time according to Eq.(22).}
\end{figure}
where $T$ is the interaction time. The instantaneous energy spectrum of the Hamiltonian $H$ versus $\theta= \pi t /T$ is shown in Fig.8(b) for a SSH chain comprising $(2N-1)=31$ sites. The zero-energy topological mode is fully localized at the left edge site of the chain for $\theta=0$ (i.t. at initial time $t=0$), and at the right edge site for $\theta=\pi$ (i.e. at final time $t=T$). Note that at $\theta= \pi/2$, i.e. at time $t=T/2$, the gap closes and the zero-energy mode touches the two bands. The behavior of the transfer probability versus the interaction time $T$ is shown in Fig.8(c). From the figure one can see that a transfer probability higher than $ \sim 90 \%$ requires an interaction time $T$ larger than $T \sim 800$, i.e. about five time longer than the interaction time required in the CTAP scheme [compare Figs.4 and 8(c)].

\section{Conclusions}
Adiabatic pumping of topologically-protected edge states has emerged in recent years as a powerful tool for robust excitation transfer, with potential applications in several areas of physics such as mesoscopic quantum transport, quantum state transfer and quantum information processing. Adiabaticity is an essential requirement in topological pumping protocols which sets a lower limit of  transfer speed. In most schemes of topological pumping, such as in Thouless pumping in crystals or quasicrystals, the adiabatic state becomes delocalized and immersed into the continuum during the adiabatic cycle, requiring extremely slow evolution to avoid nonadiabatic effects. In this work we combined the concept of topological pumping and coherent tunneling by adiabatic passage to realize excitation transfer among edge and interface states, which is more robust against nonadiabatic effects as compared to traditional Thouless pumping methods. The scheme has been illustrated by considering adiabatic passage in a system sustaining two topologically-protected edge states and one interface state, realized by interfacing two dimerized SSH chains with different topological order. We compared the transfer speed of the topological CTAP protocol with other adiabatic pumping schemes in chains of the same size, such as Thouless pumping in the Rice-Mele chain and adiabatic pumping in the SSH chain with an odd number of sites. The topological CTAP transfer scheme allows one to speed up excitation transfer as compared to the other protocols, even without introducing shortcuts to adiabaticity methods.\\ 
Further developments of the present study could be envisaged, for example one could extend the analysis considering counterdiabatic control \cite{R43,R44} or other methods of shortcuts to adiabaticity \cite{R45,R46} well developed in CTAP protocols to speed up the transfer process, or one could consider to apply CTAP topological pumping protocols to non-Hermitian crystals with enhanced selectivity of edge and interface states.\\
Our results suggest that application of CTAP to topologically-protected states sustained at the interfaces of topological materials with different topological order could provide an interesting route to speed up excitation transfer with potential applications in different physical systems of major relevance in  frontrunners quantum technologies, such as superconducting quantum systems, cold atoms and integrated photonic structures.

\appendix
\section{Derivation of the three-level CTAP equations}
In this Appendix we derive the three-level CTAP equation (13) given in the main text, which describes the approximate dynamical evolution of the system in the subspace of topological states $|L \rangle$, $|C \rangle$ and $|R \rangle$. To this aim,  we use the variational method for time-dependent problems \cite{R47,R48}. The Schr\"odinger equation 
\begin{equation}
i \frac{\partial |\psi \rangle}{\partial t}=H(t) | \psi(t) \rangle
\end{equation}
 can be derived from the variational principle 
 \begin{equation}
  \delta \int dt L(\psi, \psi^*, \psi_t, \psi_t^*,t)=0
  \end{equation}
   with the Lagrangian
   \begin{equation}
   L(\psi, \psi^*, \psi_t, \psi_t^*,t)= \frac{i}{2} \langle \psi | \psi_t \rangle-\frac{i}{2} \langle \psi_t | \psi \rangle-\langle \psi | H(t)| \psi \rangle
   \end{equation}
Let us now make the following Ansatz for the state vector
\begin{equation}
| \psi(t) \rangle \simeq  \sum_n a_n(t) |n \rangle
\end{equation}
with $n=L,C,R$. Such an Ansatz describes the approximate evolution of the state vector in the subspace of the topological states $|L \rangle$, $|C \rangle$ and $|R \rangle$. Substitution of Eq.(A4) into Eq.(A3) and from the Euler-Lagrangian equation one readily obtains the following equations for the temporal evolution of the occupation amplitudes $a_n$
\begin{equation}
i \sum_l \langle n|l \rangle \dot{a}_n=\sum_l \langle n | H(t) |l \rangle a_l -i \sum_l \langle n | \dot{l} \rangle a_l
\end{equation}
($n,l=L,C,R$), where the dot denotes the derivative with respect to time $t$. From the form of $|L \rangle$, $|C \rangle$ and $|R \rangle$ states given by Eq.(6) in the main text, it readily follows that
\begin{eqnarray}
\langle L | L \rangle=\langle C | C \rangle= \langle R  | R \rangle=1 \nonumber \\
\langle L | C \rangle=\langle R | C \rangle=0 \nonumber \\
\langle L | \dot{L} \rangle=\langle C | \dot{C} \rangle=\langle R | \dot{R} \rangle=0 \nonumber \\
\langle L | \dot{C} \rangle=\langle \dot{L} | C \rangle=\langle R | \dot{C} \rangle=\langle \dot{R} | C \rangle=0 \nonumber \\
\langle L |H|L \rangle=\langle R |H|R \rangle=\langle C |H|C \rangle=\langle L |H|R \rangle=0. \nonumber
\end{eqnarray}
Then Eq.(A5) takes the simplified form
\begin{eqnarray}
i \dot{a}_L+i \langle L | R \rangle \dot{a}_R =  \langle L | H |C \rangle a_C-i \langle L | \dot{R} \rangle a_R \\
i \dot{a}_C= \langle C | H| L \rangle a_L+\langle C | H | R \rangle a_R \\
i \dot{a}_R+i \langle R | L \rangle \dot{a}_L =  \langle R | H |C \rangle a_C-i  \langle R | \dot{L} \rangle a_L.
\end{eqnarray}
Since the states $|L \rangle$ and $|R \rangle$ are localized at the left and right edge sites, for a sufficiently long chain one has $ | \langle L | R \rangle  | \ll  1$, and hence one can set $\langle L | R \rangle=\langle R | L \rangle \simeq 0$ on the left hand side of Eqs.(A6) and (A8). Similarly, for a slow evolution of the Hamiltonian $H(t)$ the terms $\langle L | \dot{R} \rangle$ and  $\langle R | \dot{L} \rangle$ are much smaller than 
$\langle L |H |C \rangle$ and $\langle R |H |C \rangle$, and thus they can be neglected on the right hand sides of Eqs.(A6) and (A8). Under such approximations, one obtains
\begin{eqnarray}
i \dot{a}_L & \simeq & \langle L | H |C \rangle a_C \\
i \dot{a}_C & = &  \langle C | H| L \rangle a_L+\langle C | H | R \rangle a_R \\
i \dot{a}_R & \simeq  & \langle R | H |C \rangle a_C
\end{eqnarray}
which are Eqs.(13) given in the main text. The scalar products $\Omega_L \equiv \langle L | H | C \rangle$ and $\Omega_R \equiv \langle R | H | C \rangle$ can be readily calculated from Eqs.(2-6) given in the main text, yielding the final result expressed by Eqs.(14) and (15).


\begin{thebibliography}{31}

\bibitem{R1}
D.J. Thouless, Phys. Rev. B {\bf 27}, 6083 (1983).
\bibitem{R2}
J. E. Avron, D. Osadchy, and R. Seiler, Phys. Today {\bf 56}, 38 (2003).
\bibitem{R3}
D. Xiao, M.-C. Chang, and Q. Niu, Rev. Mod. Phys. {\bf 82}, 1959 (2010).
\bibitem{R3bis}
J. K. Asb\'oth, L. Oroszl\'any, and A. P\'alyi, {\it A Short Course on
Topological Insulators}, Lecture Notes in Physics (2016), Vol. 919, chap. 4.
\bibitem{R4}
M. Switkes, C.M. Marcus, K. Campman,  and A.C. Gossard, Science {\bf 283}, 1905 (1999).
\bibitem{R5}
M. D. Blumenthal, B. Kaestner, L. Li, S. Giblin, T. J. B. M.
Janssen, M. Pepper, D. Anderson, G. Jones, and D. A.
Ritchie, Nat. Phys. {\bf 3}, 343 (2007).
\bibitem{R6}
H.-I Lu, M. Schemmer, L.M. Aycock, D. Genkina, S. Sugawa, and I.B. Spielman,
Phys. Rev. Lett. {\bf 116}, 200402 (2016).
\bibitem{R7}
S. Nakajima, T. Tomita, S. Taie, T. Ichinose, H. Ozawa, L.
Wang, M. Troyer, and Y. Takahashi, Nat. Phys. {\bf 12}, 296 (2016).
\bibitem{R8}
M. Lohse, C. Schweizer, O. Zilberberg, M. Aidelsburger,
and I. Bloch, Nat. Phys. {\bf 12}, 350 (2016).
\bibitem{R9}
M. Lohse, C. Schweizer, H.M. Price, O. Zilberberg, and I. Bloch, Nature {\bf 553} 55 (2018).
\bibitem{R10}
Y. E. Kraus, Y. Lahini, Z. Ringel, M. Verbin, and O. Zilberberg,
Phys. Rev. Lett. {\bf 109}, 106402 (2012).
\bibitem{R11}
M. Verbin, O. Zilberberg, Y. Lahini, Y. E. Kraus, and
Y. Silberberg,Phys. Rev. B. {\bf 91}, 064201 (2015).
\bibitem{R12}
O. Zilberberg, S. Huang, J. Guglielmon, M. Wang, K. P.
Chen, Y. E. Kraus, and M. C. Rechtsman, Nature {\bf 553},
59 (2018).
\bibitem{R13}
M.D. Schroer, M.H. Kolodrubetz, W.F. Kindel, M. Sandberg, J. Gao, M.R. Vissers, D.P. Pappas, A. Polkovnikov, and K.W. Lehnert,
Phys. Rev. Lett. {\bf 113}, 050402 (2014).
\bibitem{R14}
W. Ma, L. Zhou, Q. Zhang, M. Li, C. Cheng, J. Geng, X. Rong, F. Shi, J. Gong, and J. Du,
Phys. Rev. Lett. {\bf 120}, 120501 (2018).
\bibitem{R15}
X. Gu, S. Chen, and Y.-x. Liu, arXiv:1711.06829 (2017).
\bibitem{R16}
R. Wang, X. Z. Zhang, and Z. Song, Phys. Rev. A {\bf 98}, 042120 (2018).
\bibitem{R18}
N.Y. Yao, C.R. Laumann, A.V. Gorshkov, H. Weimer, L. Jiang, J.I. Cirac, P. Zoller, and M.D. Lukin,
Nat. Commun. {\bf 4}, 1585 (2013).
\bibitem{R19}
S. Ganeshan, K. Sun, and S. Das Sarma, Phys. Rev. Lett. {\bf 110}, 180403 (2013).
\bibitem{R20}
N. Lang and H. P. B\"uchler, npj Quant. Inform. {\bf 3}, 47 (2017),
\bibitem{R21}
F. Mei, G. Chen, L. Tian, S.-L. Zhu, and S. Jia, 
Phys. Rev. A {\bf 98}, 012331 (2018).
\bibitem{R22}
S. Longhi, G.L. Giorgi, and R. Zambrini, Adv. Quant. Technol., 1800090 (2019). 
\bibitem{referee1}
W. P. Su,  J.R. Schrieffer, and A.J. Heeger, Phys. Rev. Lett. {\bf 42}, 1698 (1979).
\bibitem{referee2}
W.P. Su and J.R. Schrieffer, PNAS {\bf 77}, 5626 (1980). 
\bibitem{referee3}
A.J. Heeger, Phil. Trans. R. Soc. Lond. A {\bf 314}, 17 (1985).
\bibitem{referee4}
M. Kuwabara, Y. Ono, and A. Terai, J. Phys. Soc. Jpn. {\bf 60}, 1286 (1991).
\bibitem{referee5}
H. Takayama, Y. R. Lin-Liu, and K. Maki, Phys. Rev. B {\bf 21}, 2388 (1980).
\bibitem{R23}
K. Eckert, M. Lewenstein, R. Corbalan, G. Birkl, W. Ertmer, and
J. Mompart, Phys. Rev. A {\bf 70}, 023606 (2004).
\bibitem{R24}
A. D. Greentree, J. H. Cole, A. R. Hamilton, and L. C. L. Hollenberg,
Phys. Rev. B {\bf 70}, 235317 (2004).
\bibitem{R25}
K. Eckert, J. Mompart, R. Corbalan, M. Lewenstein, and G. Birkl, Opt. Commun. {\bf 264}, 264 (2006).
\bibitem{R26}
S. Longhi, G. Della Valle, M. Ornigotti, and P. Laporta, Phys. Rev. B {\bf 76}, 201101 (2007); 
G. Della Valle, M. Ornigotti, T. Toney Fernandez, P. Laporta, S. Longhi, A. Coppa, and V. Foglietti, Appl. Phys. Lett. {\bf 92}, 011106 (2008).
\bibitem{R27}
J. H. Cole, A. D. Greentree, L. C. L. Hollenberg, and S. Das Sarma, Phys. Rev. B {\bf 77}, 235418 (2008).
\bibitem{R27bis}
F. Dreisow, A. Szameit, M. Heinrich, R. Keil, S. Nolte, A. T\"unnermann, and S. Longhi, Opt. Lett. {\bf 34}, 2405 (2009).
\bibitem{R28} 
R. Menchon-Enrich, A. Benseny, V. Ahunger, A. D.
Greentree, T. Busch, and J. Mompart, Rep. Prog. Phys. {\bf 79}, 074401 (2016).
\bibitem{R29}
E.M. Graefe, H.J. Korsch, and D. Witthaut, Phys. Rev. A {\bf 73}, 013617 (2006).
\bibitem{R30}
K. Eckert, O. Romero-Isart, and A. Sanpera, New J. Phys. {\bf 9}, 155 (2007).
\bibitem{R30a}
L.M. Jong, A.D. Greentree, V.I. Conrad, L.C.L. Hollenberg, and D.N. Jamieson, Nanotechnology {\bf 20}, 405402 (2009).
\bibitem{R30b}
R. Menchon-Enrich, A. Llobera, J. Vila-Planas, V.J. Cadarso, J. Mompart, and V. Ahufinger,  Light: Science \& Appl. {\bf 2}, e90  (2013)
\bibitem{R30c}
R. Menchon-Enrich, J. Mompart, and V. Ahufinger,  Phys. Rev. B {\bf 89}, 094304 (2014).
\bibitem{R31}
N.V. Vitanov, M. Fleischhauer, B.W. Shore, and K Bergmann,
Adv. At. Mol. Opt. Phys. {\bf 46}, 55 (2001).
\bibitem{R31b}
K. Bergmann, N.V. Vitanov, and   B.W. Shore, J. Chem. Phys. {\bf 142}, 170901 (2015).
\bibitem{R32}
N.V. Vitanov, A.A. Rangelov, B.W. Shore, and K. Bergmann,
Rev. Mod. Phys. {\bf 89}, 015006 (2017).
\bibitem{R33}
L. P. Yatsenko, S. Guerin, and H. R. Jauslin,
Phys. Rev. A {\bf 65}, 043407 (2002).
\bibitem{R34}
B. W. Shore, K. Bergmann, J. Oreg, and S. Rosenwaks,
Phys. Rev. A {\bf 44}, 7442 (1991).
\bibitem{R35}
S. Longhi, Phys. Lett. A {\bf 359}, 366 (2006).
\bibitem{R36}
L. Privitera, A. Russomanno, R. Citro, and G.E. Santoro, Phys. Rev. Lett. {\bf 120}, 106601 (2018).
\bibitem{R37}
C. Poli, M. Bellec, U. Kuhl, F. Mortessagne, and H. Schomerus,
Nat. Commun. {\bf 6}, 6710 (2015).
\bibitem{R38}
Indeed, adiabatic theorems without a gap condition can be stated. See, for instance:
J.E. Avron and A. Elgart, Commun. Math. Phys. {\bf 203}, 445 (1999).
\bibitem{R39}
M. Bello, C. E. Creffield, and G. Platero, Sci. Rep.  {\bf 6}, 22562 (2006).
\bibitem{R40}
G.M.A. Almeida, F. Ciccarello, T.J.G. Apollaro, and A.M.C. Souza, Phys. Rev. A {\bf 93}, 032310 (2016).
\bibitem{R41}
N. Lang and H.P.B\"uchler, npj Quantum Inf. {\bf 3}, 47 (2017).
\bibitem{R42}
M. P. Estarellas, I. D$^{\prime}$Amico, and T. P. Spiller, Sci. Rep. {\bf 7},  42904 (2007).
\bibitem{R43}
 M. Demirplak and S. A. Rice, J. Phys. Chem. A {\bf 107}, 9937 (2003).
 \bibitem{R44}
 M. V. Berry, J. Phys. A {\bf 42}, 365303 (2009).
 \bibitem{R45}
 S. Ib\'a\~{n}ez, X. Chen, E. Torrontegui, J. G. Muga, and A.
Ruschhaupt, Phys. Rev. Lett. {\bf 109}, 100403 (2012).
\bibitem{R46}
A. Baksic, H. Ribeiro, and A. A. Clerk, Phys. Rev. Lett. {\bf 116}, 230503
(2016).
\bibitem{R47}
P. Kramer and M. Saraceno, {\it Geometry of the Time-Dependent Variational Principle
in Quantum Mechanics} (Springer-Verlag, Berlin) (1981).
\bibitem{R48}
P. Kramer, J. Phys.: Conf. Ser. {\bf 99}, 012009 (2008).





 



\end{thebibliography}

\end{document}